# Adaptive beamforming for optical wireless communication via fiber modal control

Chao Li[1,†], Yiwen Zhang[2,†], Xinda Yan[3,†], Yuzhe Wang[3], Xuebing Zhang[3], Jian Cui[4], Lei Zhu[3], Juhao Li[4], Zilun Li[5], Shaohua Yu[1], Zizheng Cao[1,*], A.M.J. Koonen[3] and Chia Wei Hsu[2,*]

[1]Peng Cheng Laboratory, Shenzhen 518055, Guangdong, China.
[2]Ming Hsieh Department of Electrical and Computer Engineering, University of Southern California, Los Angeles, California 90089, USA.
[3]Institute for Photonic Integration, Department of Electrical Engineering, Eindhoven University of Technology, Eindhoven, 5612 AZ, The Netherlands.
[4]State Key Laboratory of Advanced Optical Communication Systems and Networks, Peking University, Beijing 100871, China.
[5]Division of Vascular Surgery, The First Affiliated Hospital, Sun Yat-sen University, Guangzhou 510080, Guangdong, China.

*Corresponding author(s). E-mail(s): caozzh@pcl.ac.cn; cwhsu@usc.edu;
†These authors contributed equally to this work.

**Abstract**

High-speed optical wireless communication can address the exponential growth in data traffic. Adaptive beamforming customized for the target location is crucial, but existing solutions such as liquidcrystal spatial light modulators and microelectromechanical systems require costly micro/nano manufacturing, delicate alignment, and a high degree of mechanical stability. These challenges reflect the fragility of integrating a fiber network with micro/nano mechanical or photonic systems. Here, we realize low-cost, low-loss, and fiber-compatible beamforming and continuous beam steering through controlled bending of a multi-mode fiber that modifies its modal coupling, and use it to enable flexible optical wireless communication at 10 Gb/s. By using the fiber modal coupling as degrees of freedom rather than an impediment, this approach offers a promising solution for flexible and cost-effective optical wireless communication networks.

# 1 Introduction

The spectral efficiency of wireless communication has increased 1 million fold over the last 45 years [1], and continued advance is necessary to keep up with the continued growth of data traffic coming from remote work, home entertainment, smart buildings, AR/VR devices, etc. Higher radio frequencies in the millimeter-wave band can realize larger bandwidth [2, 3], and optical wireless communication (OWC) [4–6] uses higher electromagnetic frequencies in the ultraviolet, visible, and infrared bands to achieve even faster speed and higher capacity. OWC has been identified as a promising solution to the radio frequency spectrum issues crisis [7, 8], especially for indoor communication which accounts for the majority of mobile data traffic [1]. Global standardization activities within IEEE 802.11bb have been carried out to determine the technical and economic opportunities of OWC [9].

Due to the link budget constraints [10] including safety restrictions [11] and noise from ambient light [12], high-bit-rate optical communication requires the wireless signal to be focused towards the target receiver rather than broadcast to all directions. In a typical indoor OWC scenario (Fig. 1a), adaptive beamforming transmitters [4, 5, 13–15] realize steerable beams towards users that can move in the coverage area. A practical OWC system requires the beamforming transmitters to be low-cost, low-loss, and easy to assemble. To date, several approaches have been proposed to realize beamforming OWC systems. Systems with actively controlled elements like liquid-crystal spatial light modulators (LC-SLMs) [16–18] or microelectromechanical systems (MEMS) [19] offer high flexibility but are expensive and require careful alignment to couple light from the network fiber to the free-space optical element; LC-SLM also suffers from diffraction loss and polarization dependence. Using wavelength-dependent diffraction, the architecture based on diffraction grating [20] is a simple solution but also suffers from diffraction loss and polarization dependence. Arrayed waveguide grating router (AWGR) [21] can be directly connected to an existing fiber network, but it comes with large coupling loss and can only steer the beam to discrete directions. All solutions proposed so far require costly micro/nano manufacturing, delicate assembling, and a high degree of mechanical stability to maintain the coupling efficiency These challenges come from the fundamental difficulty of integrating fiber technology into micro/nano mechanical and photonic systems.

Here we propose an all-fiber-based adaptive beamforming transmitter with flexible steering capability and high coupling efficiency. As illustrated in Fig. 1b, the proposed fiber-based spatial light modulator (F-SLM) uses a multimode fiber (MMF) that can directly connect to the optical fiber network with a low coupling loss and low manufacturing cost [22, 23]. Instead of modulating the optical wavefront with spatial pixels, F-SLM controls the modal interactions among propagating modes in the MMF [24] by wrapping the MMF around several paddles and optimizing the rotation angle of the paddles. The tailored modal coupling leads to the optimized superposition of fiber modes and shapes the outgoing optical wavefront. This approach enables beamforming with a continuous angular scan while avoiding free-space coupling loss, polarization dependence, and diffraction loss. As the F-SLM does not require any delicate micro/nano device, it provides a robust and cost-effective solution for beamforming OWC systems. Using the F-SLM-based beamforming transmitter, we achieve a 10-Gb/s OWC system with a continuous 2D beam steering.

## 2 F-SLM beamforming transmitter

The MMF of the F-SLM is connected to a section of single-mode fiber which delivers the modulated light source and can come directly from the fiber access network to the building [1, 23]. In F-SLM, the MMF wraps around several motorized paddles (Thorlabs, MPC320) (Fig. 1b) to create controlled bending. Such bending leads to coupling between the two polarization states of each fiber mode, commonly used for polarization control in a single-mode fiber [25, 26], as well as coupling between different fiber modes [27–29]. By rotating the paddles, we modulate the polarization coupling and spatial-mode coupling in the MMF to control the wavefront of the outgoing light. In our proof-of-concept experiment, we use a 1550-nm laser diode (Santec, MLS-2100) as the light source and a standard OM1 62.5/125 μm graded-index MMF (Thorlabs, GIF625, 0.275 NA, 5 meters long) in the F-SLM setup. The MMF is coupled to the single-mode fiber of the laser with a 15-μm offset to promote mode diversity (see Supplementary Section I and Supplementary Fig. S1). Additionally, our F-SLM setup utilized nine paddles with individual motorized controllers, each with a diameter of 18 mm, which produce a gentle bending with negligible bending loss compared to actuators that locally bend/compress the fiber [24]. A collimator ($f$ = 18 mm) and a plano-convex lens ($f$ = 40 mm) are placed after the MMF to set the divergence angle at around 5°. We refer to this system as an F-SLM-based transmitter (F-SLM Tx), which can realize adaptive beamforming toward different directions by adjusting the paddles electrically.

To characterize the beamforming performance, we first use a photodiode power sensor (Thorlabs, S122C, 9.5-mm sensor diameter) as the target receiver, placed on a receiving plane after $d$ = 80 cm of free space propagation. The paddle rotation angles in the F-SLM are optimized in a feedback loop to enhance the received optical power. We define the enhancement factor as the ratio between the optimized power and the power averaged over 120 random paddle configurations. Fig. 1c shows the two-dimensional (2D) distribution of the enhancement achieved by the F-SLM Tx (see Supplementary Fig. S2a for the optimized power and averaged power). The enhancement values are measured at 31 locations on the receiving plane, with polar coordinates ($R$, $\varphi$) of $R$ = 0, 1.4 cm, 2.8 cm, 4.2 cm, 5.6 cm and 7 cm, and $\varphi$ = 0, π/3, 2π/3, π, 4π/3, and 5π/3, covering a 2D beamsteering angular range $\eta$ = atan($R/d$) from −5° to 5°. The average enhancement over all receiving locations within this 2D 10° angular range is 5 (about 7 dB). After optimizing the F-SLM, light is focused toward the target directions as shown by CCD (Goldeye, CL-008) images in Fig. 1d; before the optimization, light is spread over a wide range (see Supplementary Fig. S2b).

## 3 Wireless communication with F-SLM

We use the F-SLM Tx to realize a flexible 10-Gb/s OWC system, with the experimental setup shown in Fig. 2a and details described in Supplementary Section III. A bit error ratio (BER) tester drives a Mach–Zehnder modulator with a 10-Gbaud on-off keying (OOK) signal at 1 V peak-to-peak voltage. A laser diode produces light at a wavelength of 1550 nm, which is modulated using a Mach-Zehnder modulator, amplified with an erbium-doped fiber amplifier, and then transmitted through the F-SLM before being emitted into free space. After traveling 80 cm through free space, the modulated optical signal from the F-SLM is coupled into a single-mode fiber using a fiber collimator and MMF. This signal is then converted into an electrical signal via an avalanche photodiode (APD) and fed back into the BER tester for analysis (see Supplementary Fig. S3a for a picture of the experimental setup). All BER values are measured with 60 seconds of accumulation time by the BER tester.

Figure 2b shows the optical powers received by the APD and the corresponding BER values before and after optimizing the F-SLM. Without optimization, the average received optical power ranges between −41.8 dBm and −32 dBm at different angles (as indicated by the black curve with triangle marks in Fig. 2b). These values fall below the sensitivity threshold of the APD, so the electrical signal cannot be demodulated, and the BER remains at approximately 0.5 (represented by the black line with circular marks in Fig. 2b). After the F-SLM optimization, the optical power received by the APD ranges between −28.5 dBm and −25.5 dBm, which is sufficient to be demodulated and translates to a recovered BER value between $3.6 \times 10^{-3}$ and $2 \times 10^{-11}$. Such pre-error-correction BER is sufficient for reducing the post-error-correction BER below $10^{-9}$ using forward error correction with a 7% overhead, where the pre-correction BER needs to be lower than $3.8 \times 10^{-3}$ [30].

We further characterize the BER performance as a function of the received optical power in Fig. 2c for different receiver locations after F-SLM optimization. The received powers are adjusted in steps of 0.5 dBm by a variable optical attenuator. The optical wireless link employing the F-SLM achieves a BER that is comparable to a back-to-back connection without the 80-cm free space propagating and the F-SLM, also shown in Fig. 2c. The incorporation of the F-SLM beamforming transmitter results in only a slight increase in BER. The BER performance is consistent with the measured optical spectra at the received power of -26 dBm (see Supplementary Fig. S3b), which shows almost the same optical signal-to-noise-ratio. The eye diagram of the 10 Gb/s OOK signal (with a received optical power of -26 dBm and a BER of $3.7 \times 10^{-9}$) is shown in the inset of Fig. 2c.

## 4 Theoretical modelling

To interpret the working mechanism of the F-SLM and to predict its scaling for larger systems, we develop a theoretical model, illustrated in Fig. 3a, to describe light propagation in the F-SLM. In this model, an MMF supporting $N$ fiber modes is divided into $K$ sections, and each section is wrapped around one paddle. For a graded-index MMF, the propagating modes are divided into mode groups, with modes in one group sharing a common propagation constant [31].

Each mode in the MMF has a two-fold degeneracy corresponding to two orthogonal polarization states. Thus, the light propagation in the $k$-th ($k = 1, 2, \cdots, K$) MMF section can be modelled by a $2N \times 2N$ transmission matrix $T_k$, which we further write as the product of three matrices, $M_k^{C1}$, $M_k^{J}$ and $M_k^{C2}$. The matrices $M_k^{C1}$ and $M_k^{C2}$ use random unitary matrices to model the random coupling of modes and both of their polarization states within each mode group of the graded-index MMF; cross coupling between mode groups is ignored [32, 33]. The matrix $M_k^{J}$ uses Jones matrices [34, 35] to model the additional coupling between the two polarization states through bending induced birefringence [25, 26]. As described in Supplementary Section IV, $M_k^{J}$ is a function of the polarization rotation angle $\theta_k$ which is induced by rotating the $k$-th paddle in F-SLM and can be changed from 0 to $2\pi$. Therefore, the output field of the F-SLM becomes a function of the $K$ rotation angles $\theta_1$, $\theta_2, \cdots, \theta_K$. These angles are optimized to maximize the intensity in one output speckle, summed over both polarization components (see Supplementary Section IV for details).

The initial power distribution among mode groups in the MMF determines how many modes participate. We model the single-mode-to-multi-mode excitation (see Supplementary Section IV) to estimate such power distribution (Supplementary Fig. S4), which shows that the energy is primarily distributed among mode groups 3 through 8 (including 33 modes).

Using this matrix model with the estimated initial power distribution, we investigate the enhancement of the F-SLM as a function of the number of optimized paddles for different random configurations. Figure 3b shows the predicted theoretical range of F-SLM enhancement within one standard deviation (green shading), which is consistent with the experimental results (red curve). To interpret the source of the enhancement, we remove the spatial-mode-coupling matrices ($M_k^{C1}$ and $M_k^{C2}$ in $T_k$) from the model, keeping only the polarization coupling matrix $M_k^J$; in this case, the predicted enhancement (orange shading in Fig. 3b) is substantially lower than the experimental results and stops increasing beyond 4 paddles. This comparison shows that the coupling between different fiber modes is critical for the observed enhancements.

To predict how the F-SLM performance scales, we consider a uniform excitation of $N$ fiber modes and study the F-SLM enhancement with an increasing number of paddles and increasing number $N$ of excited modes. Figure 3c shows the theoretical prediction of the average F-SLM enhancement with 6, 15, 45 and 105 excited modes and up to 15 paddles. When the number of paddles exceeds the number $N$ of excited modes, the enhancement saturates toward the maximal factor-of-$N$ enhancement. In the opposite limit where $N$ is sufficiently larger than the number of paddles, we observe a linear relationship between the enhancement and the paddle number, with a slope of 0.70. In comparison, for an LC-SLM when both polarization components are measured at one speckle in the output, the slope between enhancement and the number of LC-SLM macropixels is $0.5(\pi/4) \approx 0.39$ with phase-only modulation [36]. Therefore, each paddle in the F-SLM achieves enhancement close to that from two phase-modulation LC-SLM macropixels.

## 5 Discussions

Table 1 Comparison of beamforming OWC systems.

|  | F-SLM | LC-SLM[16] | MEMS[19] | Grating[20] | AWGR[21] |
|---|---|---|---|---|---|
| **Fiber compatibility** | ✓ | ✗ | ✗ | ✗ | ✓ |
| **Works for both polarizations** | ✓ | ✗ | ✓ | ✗ | ✗ |
| **Angular continuity** | ✓ | ✓ | ✓ | ✓ | ✗ |
| **Insertion/coupling loss** | 7% | 13% | 7% | 13% | 30% |
| **Cost** | Low | High | High | Low | Medium |
| **Scanning range** | 10°×10° | 3°×3°✚ | 20°×20° | 5.6°×12.6° | 17°×17° |

✚ 30°×30° with angle magnifier [17].

Table 1 summarizes the properties of the F-SLM beamforming transmitter in comparison to other infrared beamforming transmitters using LC-SLM [16, 17], MEMS [19], gratings [20], and AWGR [21]. F-SLM Tx is compatible with a fiber access network, has low loss, works for both polarizations, and offers a continuous beam steering angle in 2D. The scanning range of F-SLM can be further increased by exciting higher-order fiber modes and by incorporating an angle magnification module [17]. Excitation of more modes will also enable larger F-SLM enhancement when more paddles are used. The paddle optimization can be automated with servo motors, for which response times less than 5 ms are commercially available. With its simple configuration with no micro/nano components, F-SLM Tx provides a robust and cost-effective beamforming OWC system.

**Contribution.** Z. Cao conceived the idea of fiber-based spatial light modulator and its application for beamforming optical wireless communication. Z. Li and S. Yu joined the application analysis. C. Li led the experiment. C. Li, X. Yan, Y. Wang, L. Zhu and X. Zhang performed the experiment. C.W. Hsu led the theory. Y. Zhang and C.W. Hsu performed the modelling. J. Cui and J. Li performed the MMF mode analysis. C. Li, X. Yan, and Y. Zhang performed data analysis. All authors discussed the results and contributed to the manuscript preparation. C.W. Hsu, Z. Cao, and A.M.J. Koonen supervised the research.

**Funding.** This work is supported by the Major Key Project of Peng Cheng Laboratory (PCL), National Natural Science Foundation of China (62205166) and the National Science Foundation (ECCS-2146021). Computing resources are provided by the Center for Advanced Research Computing at the University of Southern California.

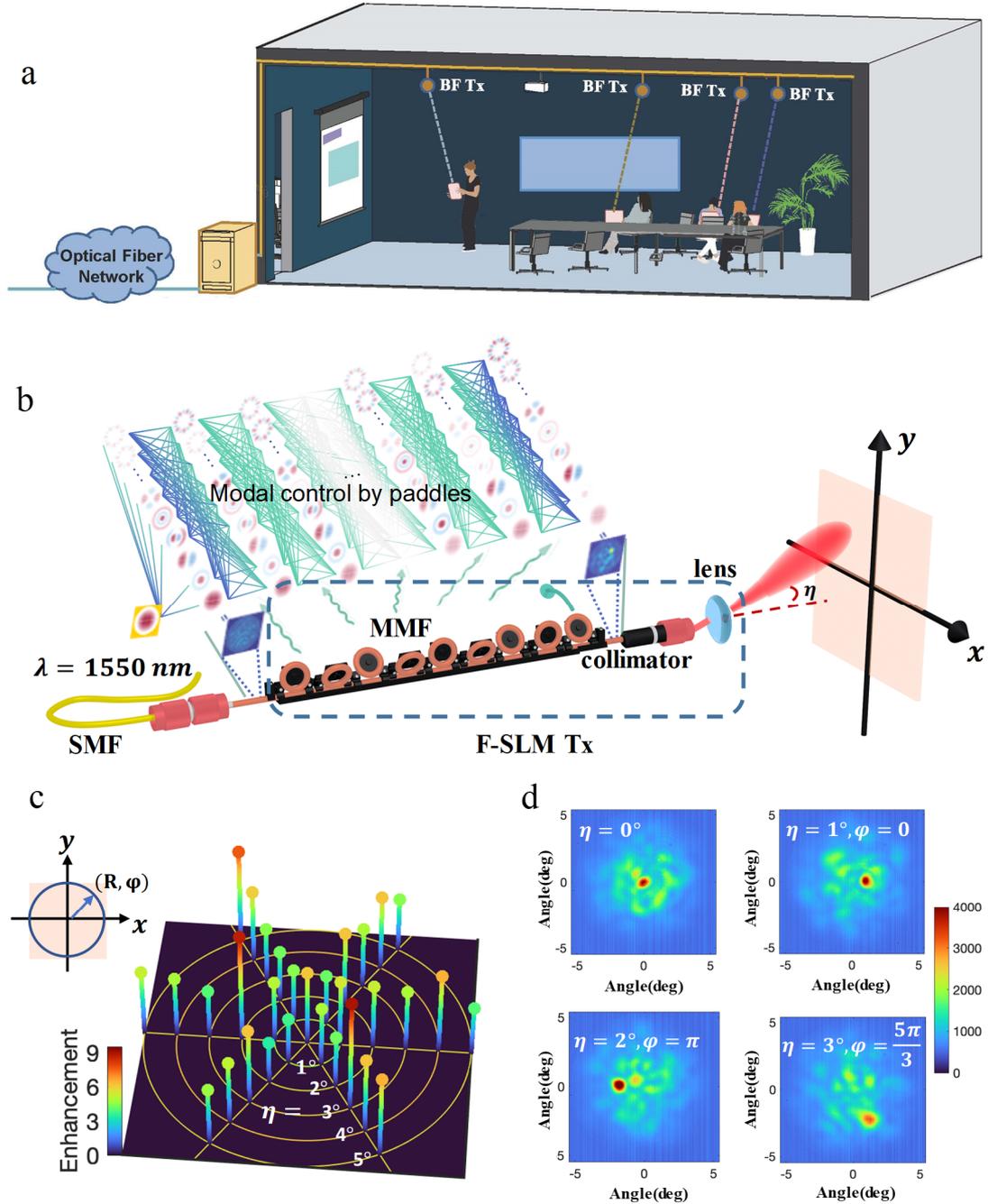

**Fig. 1** Adaptive beamforming for optical wireless communication (OWC) using a fiber-based spatial light modulator (F-SLM). (a) Schematic of an indoor infrared OWC system equipped with beamforming transmitters (BF Tx). (b) Schematic of the F-SLM: a multimode fiber (MMF) loops around paddles whose rotation angles are adjusted to control the coupling between the fiber modes, shaping the outgoing wavefront to focus toward target angle $\eta$. SMF: single-mode fiber. (c) Enhancement realized by the F-SLM at different locations on the receiving plane ($x$, $y$). (d) Intensity distribution of the outgoing light for different target angles.

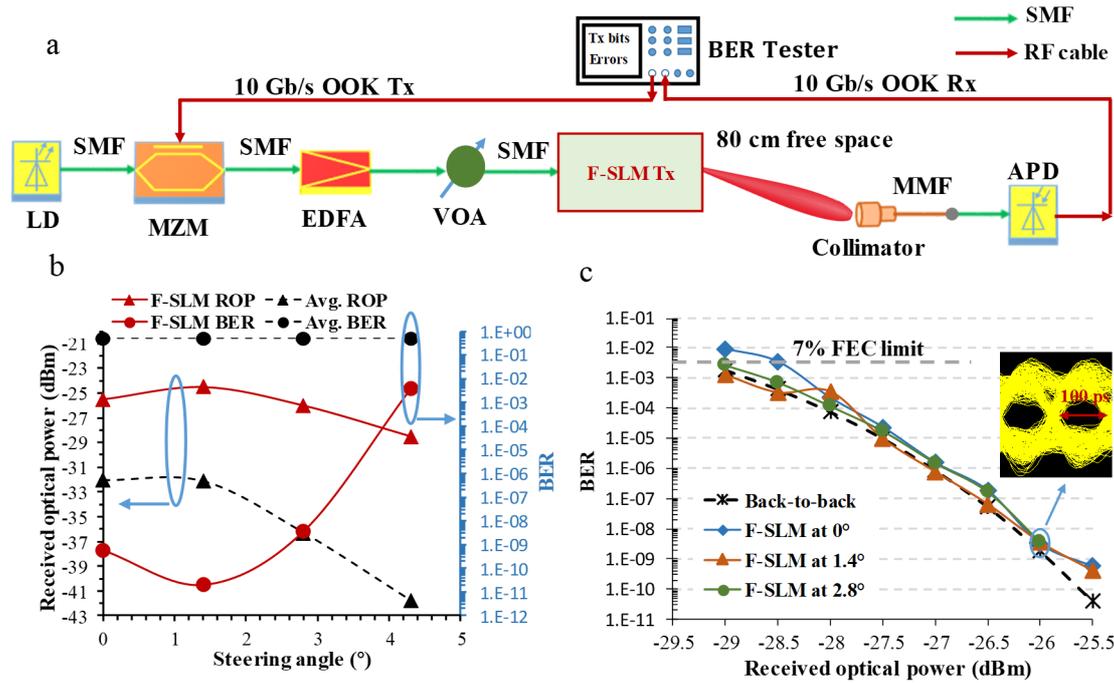

**Fig. 2** Optical wireless communication (OWC) assisted by the F-SLM. (a) Schematic setup of a 10 Gb/s on-off keying (OOK) optical wireless data transmission with beamforming realized by the F-SLM. LD: laser diode; MZM: Mach-Zehnder modulator; EDFA: Erbiumdoped fiber amplifier; VOA: variable optical attenuator; APD: avalanche photodiode. (b) Received optical power (ROP) by the APD and bit error ratio (BER) of data transmission with F-SLM (red solid lines) and without F-SLM (black dashed lines). (c) Measured BER at different steering angles as the received optical power (ROP) is tuned by the VOA, in comparison to a back-to-back link without going through free space. Inset shows an example of the eye diagram. FEC: forward error correction.

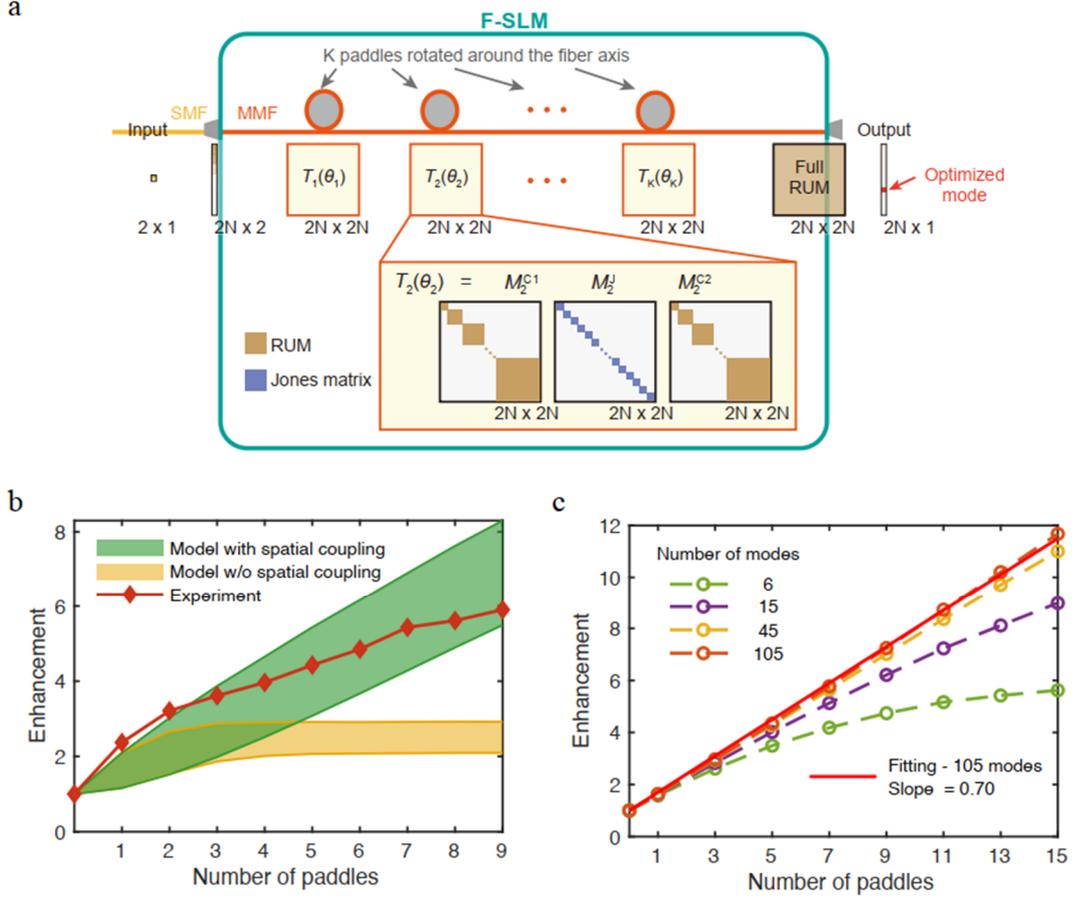

**Fig. 3** Theoretical modelling of the F-SLM. (a) Schematic of the model. Incoming light from the SMF excites $N$ spatial modes in the MMF (with a nonuniform distribution, illustrated by different colors in the excitation matrix). Matrix $T_k(\theta_k)$ is the transmission matrix of the $k$-th MMF section as a product of three matrices, $M_k^{C1}$, $M_k^J$ and $M_k^{C2}$. $M_k^{C1}$ and $M_k^{C2}$ model the coupling between spatial modes within each mode group using random unitary matrices (RUM). $M_k^J$ models the coupling between the two polarization states of each spatial mode via Jones matrices with the polarization rotation angle $\theta_k$ induced by rotating the $k$-th paddle. Another RUM models the conversion from the modal basis inside the MMF to the angular basis at the output. (b) Measured and modelled enhancement as a function of the number of optimized paddles; in the model without spatial coupling, matrices $M_k^{C1}$ and $M_k^{C2}$ are removed. Shading indicates plus/minus one standard deviation among random configurations. (c) Modelled average enhancement given different numbers of evenly-excited fiber modes.

# Supplementary Materials for

## Adaptive beamforming for optical wireless communication via fiber modal control

Chao Li[1,†], Yiwen Zhang[2,†], Xinda Yan[3,†], Yuzhe Wang[3], Xuebing Zhang[3], Jian Cui[4], Lei Zhu[3], Juhao Li[4], Zilun Li[5], Shaohua Yu[1], Zizheng Cao[1,*], A.M.J. Koonen[3] and Chia Wei Hsu[2,*]

[1]Peng Cheng Laboratory, Shenzhen 518055, Guangdong, China.
[2]Ming Hsieh Department of Electrical and Computer Engineering, University of Southern California, Los Angeles, California 90089, USA.
[3]Institute for Photonic Integration, Department of Electrical Engineering, Eindhoven University of Technology, Eindhoven, 5612 AZ, The Netherlands.
[4]State Key Laboratory of Advanced Optical Communication Systems and Networks, Peking University, Beijing 100871, China.
[5]Division of Vascular Surgery, The First Affiliated Hospital, Sun Yat-sen University, Guangzhou 510080, Guangdong, China.

*Corresponding author(s). E-mail(s): caozzh@pcl.ac.cn; cwhsu@usc.edu;
†These authors contributed equally to this work.

## I. Fiber connection

In our experiment, the multi-mode fiber (MMF) from the fiber-based spatial light modulator (F-SLM) is connected to a single-mode fiber (SMF) (G.652) from the input laser source. Both the MMF and SMF are assembled with FC/PC connectors and mounted on two separate two-dimensional translation stages, with an offset [1] between the two end faces. The power loss was measured at different offset locations to determine the fiber center (Fig. S1). We then adjusted the translation stage to generate a 15-μm offset launch with a power loss of 0.3 dB, corresponding to the 7% coupling/insertion loss in Table 1 of the main text.

## II. F-SLM characterization

Figure S2 provides extended data for the F-SLM characterization described in the main text.

## III. OWC experiment

In our optical wireless communication (OWC) experiment, the transmitted optical carrier is generated by a laser diode (Santec, MLS-2100) with 10 dBm output optical power and 1550 nm wavelength and then modulated by a 10 Gb/s OOK signal from a bit-error-ratio (BER) tester (Luceo, EPG10) via a Mach-Zehnder modulator (MZM, Fujitsu, FTM7937EZ). The optical power after the MZM is about 0 dBm, and we use an Erbium-doped fiber amplifier (Amonics, AEDFA-23-B-FA) to amplify the optical power to around 10 dBm and launch it into the F-SLM.

Figure S3a shows a picture of the F-SLM transmitter with nine motorized paddles in the OWC link. At the receiver, after 80 cm of free-space transmission, the received optical beam is coupled into a short section of multi-mode fiber (MMF) via a fiber collimator (Thorlabs, F810FC-1550) and then fed into the SMF input of an avalanche diode (APD) (Oclaro, AT10GC) for optical-to-electrical conversion. The output electrical OOK signal from the APD is then fed back into the receiver side of the BER tester. No optical amplifier is used at the receiver. We measure the optical spectra at the received optical power of -26 dBm for a back-to-back link and for F-SLM, shown in Fig. S3b, which indicate almost the same value of optical signal-to-noise ratio.

## IV. Theoretical modelling of F-SLM

We use a matrix model, illustrated in Fig. 3a of the main text, to describe light transmission through an MMF wrapped around several paddles. In graded-index MMF, the propagating modes are divided into mode groups, with modes in one group sharing a common propagation constant [2, 3]. We consider an MMF with $P$ mode groups, indexed by the integer $p = 1, 2, \ldots, P$. The $p$-th mode group has $n_p$ modes, and the total number of modes in the MMF is $N = \sum_{p=1}^{P} n_p$. As each mode has a two-fold degeneracy corresponding to the two orthogonal polarization states, light propagation in the MMF can be modelled as a $2N \times 2N$ matrix. When light propagates through the MMF, bending, index variations, and other perturbations allow power to couple from one mode to another. The coupling between modes in different groups is weak [4, 5], while the coupling between modes in the same group is strong [6–8]. For simplicity, we assume that inter-group mode coupling is negligible and that all modes and both of their polarization states within a group are fully coupled in a random manner. With this assumption, we use a $2n_p \times 2n_p$ random unitary matrix $U_p$ to model the mode and polarization coupling within the $p$-th mode group. The mode coupling considering all the groups is then modelled by a $2N \times 2N$ block diagonal matrix with independent random unitary matrices aligned along the main diagonal, written as

$$M^C = \mathrm{diag}[U_1, U_2, U_3, \cdots, U_P]. \tag{S1}$$

We divide the MMF into $K$ sections, each of which loops around one paddle and is indexed by an integer $k = 1, 2, \cdots, K$. The bending around the paddles produces birefringence in the fiber, creating a phase retardation between the two polarization states [9]. When the paddle rotates about the fiber axis, the principle axes of the fiber rotates together, producing a polarization rotation [10, 11]. We use a Jones matrix [12] and rotation matrices to model the birefringence produced by the $k$-th paddle:

$$U_k^J(\theta_k) = \begin{pmatrix} \cos\theta_k & -\sin\theta_k \\ \sin\theta_k & \cos\theta_k \end{pmatrix} \begin{pmatrix} 1 & 0 \\ 0 & e^{i\delta_k} \end{pmatrix} \begin{pmatrix} \cos\theta_k & \sin\theta_k \\ -\sin\theta_k & \cos\theta_k \end{pmatrix} \tag{S2}$$

where $\delta_k$ and $\theta_k$ are the polarization retardation and polarization rotation angles induced by the $k$-th paddle. For simplicity, here we fix the polarization retardation angle to be $\delta_k = \pi/2$ for all paddles and for all fiber modes, and let all modes have the same polarization rotation angle $\theta_k$ induced by rotating the $k$-th paddle. The polarization mixing for all modes is then modelled by a block diagonal matrix with $N$ identical Jones matrices, namely

$$M_k^J(\theta_k) = I_N \otimes U_k^J(\theta_k) \tag{S3}$$

To model both the random coupling and the bending-induced polarization mixing, we express the transmission matrix in the $k$-th MMF section as

$$T_k(\theta_k) = M_k^{C2} M_k^J(\theta_k) M_k^{C1}, \tag{S4}$$

where $M_k^{C1}$ and $M_k^{C2}$ are the same type of block diagonal matrices as $M^C$ in Eq. (S1) to model the random mode coupling. By cascading all MMF sections, the output field is

$$E_{\text{out}}(\theta_1, \theta_2, \cdots, \theta_K) = U_{\text{out}}\left(\prod_{k=1}^{K} T_k(\theta_k)\right) U_{\text{in}} E_{\text{in}}. \tag{S5}$$

where $E_{\text{in}}$ is the input field, and $U_{\text{in}}$ and $U_{\text{out}}$ describe the coupling into and out of the MMF. Since we use a SMF as the input for F-SLM, $E_{\text{in}}$ is a 2×1 column vector (for the two polarization states of the SMF) and $U_{\text{in}}$ is a $2N \times 2$ matrix. If all modes in the MMF are randomly excited uniformly by the single-mode input, $U_{\text{in}}$ can be expressed as a submatrix of a $2N \times 2N$ random unitary matrix. But the input excitation is typically highly nonuniform, so we multiply the random unitary matrix with a weight distribution to generate $U_{\text{in}}$ (see Supplementary Section IV). Since we are interested in $E_{\text{out}}$ in the angular basis instead of fiber-mode basis, we use a $2N \times 2N$ random unitary matrix as $U_{\text{out}}$ to describe such basis change. This yields $E_{\text{out}}$ as a $2N \times 1$ column vector $[u_1^H, u_1^V, u_2^H, u_2^V, \cdots, u_N^H, u_N^V]^T$, where $u_m^H$ and $u_m^V$ are the horizontal and vertical polarization components of the $m$-th output speckle. The intensity of the $m$-th speckle including both polarization components is then $I_m(\theta_1, \theta_2, \cdots, \theta_K) = |u_m^H|^2 + |u_m^V|^2$.

To mimic the experimental condition, we optimize the $K$ variables $\theta_1, \theta_2, \cdots, \theta_K$ sequentially. When optimizing $\theta_K$, Eq. (S5) can be treated as a single-variable function $E_{\text{out}}(\theta_K) = W_k M_k^J(\theta_k) V_k$, where $V_k = M_k^{C1}\left(\prod_{i=1}^{k-1} T_i(\theta_i)\right) U_{\text{in}} E_{\text{in}}$ is a $1 \times 2N$ column vector, and $W_k = U_{\text{out}}\left(\prod_{i=k+1}^{K} T_i(\theta_i)\right) M_k^{C2}$ is a $2 \times 2N$ matrix. This allows us to reduce $I_m(\theta_k)$ to a simple sinusoidal function

$$\begin{aligned} I_m(\theta_k) &= \sum_{j=1,2} \left| a_j \cos^2\theta_k + b_j \sin^2\theta_k + c_j \cos\theta_k \sin\theta_k \right|^2 \\ a_j &= \sum_{i=0}^{N-1} w_{j,2i+1} v_{2i+1} + e^{i\delta_k} w_{j,2i+2} v_{2i+2} \\ b_j &= \sum_{i=0}^{N-1} e^{i\delta_k} w_{j,2i+1} v_{2i+1} + w_{j,2i+2} v_{2i+2} \\ c_j &= (1 - e^{i\delta_k}) \sum_{i=0}^{N-1} w_{j,2i+1} v_{2i+1} + w_{j,2i+2} v_{2i+2} \end{aligned} \tag{S6}$$

where $v$ and $w$ are elements of $V_k$ and $W_k$. We use the gradient-based interior-point algorithm implemented in the fmincon function in Matlab to maximize Eq. (S6) for each $\theta_k$, and cycle through $\theta_1, \theta_2, \cdots, \theta_K$ until an optimization step changes the intensity function value by less than 1% of the averaged intensity. To obtain statistical information, we run the optimization for 500 realizations of MMF with different random matrices.

## V. Mode group and excitation power distribution analysis

We theoretically analyze the mode groups in the MMF. The refractive index profile of the graded-index MMF we use is

$$n(r) = \begin{cases} n_1[1 - 2\Delta(\frac{r}{a})^2]^{1/2}, & r < a \\ n_2 = n_1(1 - 2\Delta)^{1/2}, & r \geq a \end{cases} \tag{S7}$$

Here, the core radius is $a = 62.5/2$ μm, the core index is $n_1 = 1.49$, and the numerical aperture is NA$= \sqrt{n_1^2 - n_2^2} = \sqrt{2\Delta} n_1 = 0.275$. For such a graded-index MMF, guided modes for each of the two polarizations can be mathematically expressed as Hermite-Gaussian (HG) functions [2, 3].

$$E_{mn}(x, y, z) = C_{mn} e^{-\frac{r^2}{2s^2}} H_m\left(\frac{x}{s}\right) H_n\left(\frac{y}{s}\right) e^{iz\beta_{mn}} \quad (S8)$$

where

$$s = \left(\frac{a^2}{2\Delta n_1^2 k_0^2}\right)^{1/4} \quad (S9)$$

$C_{mn}$ is the mode amplitude to be determined, $H_n$ is the $n$-th Hermite polynomial, $k_0 = 2\pi/\lambda$ is the vacuum wave number, $\lambda$ the wavelength, and $m$, $n = 0, 1, 2\cdots$ are the mode indices. Their propagation constants are

$$\beta_{mn} = n_1 k_0 \left[1 - \frac{4\Delta}{V}(m + n + 1)\right]^{\frac{1}{2}}. \quad (S10)$$

where $V = k_0 aNA$ is the normalized frequency. HG modes with the same $p \equiv m + n + 1$ values have the same propagation constant; they consist the $p$-th mode group with $p = 1, 2,\cdots, P$ where $P \approx V/2$ corresponds to the highest-order mode group and $\beta_{mn} \approx n_2 k_0$. In the $p$-th mode group, there are $p$ HG modes per polarization. The total number of HG modes is $N \approx P^2/2 \approx V^2/8$ per polarization.

Figure S4 shows the power distribution among the mode groups given a 15-μm offset launch, simulated using the waveguide module of Rsoft photonics component design suite. This power distribution is used to build the input-coupling matrix $U_\text{in}$ in the theoretical model of F-SLM.

(1982).

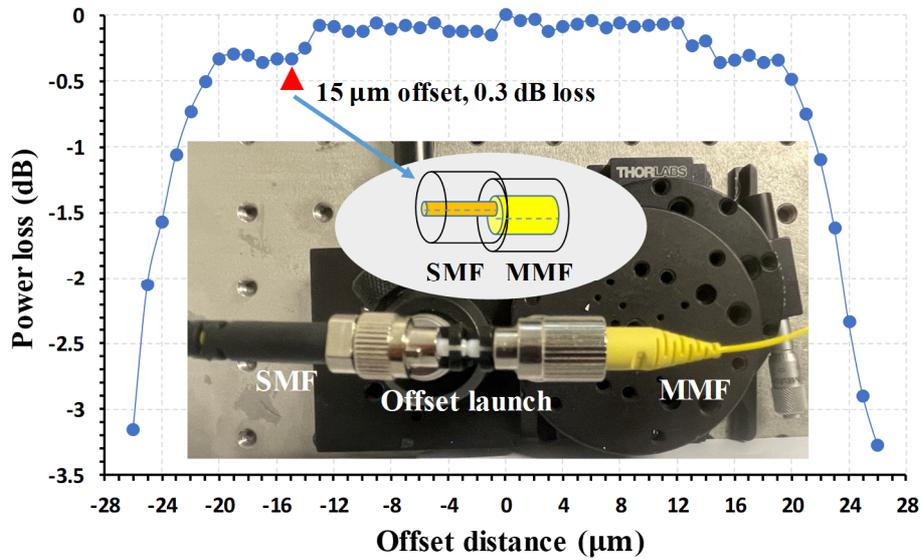

**Fig. S1** The measured power loss as a function of the offset distance between the single-mode fiber and the multi-mode fiber.

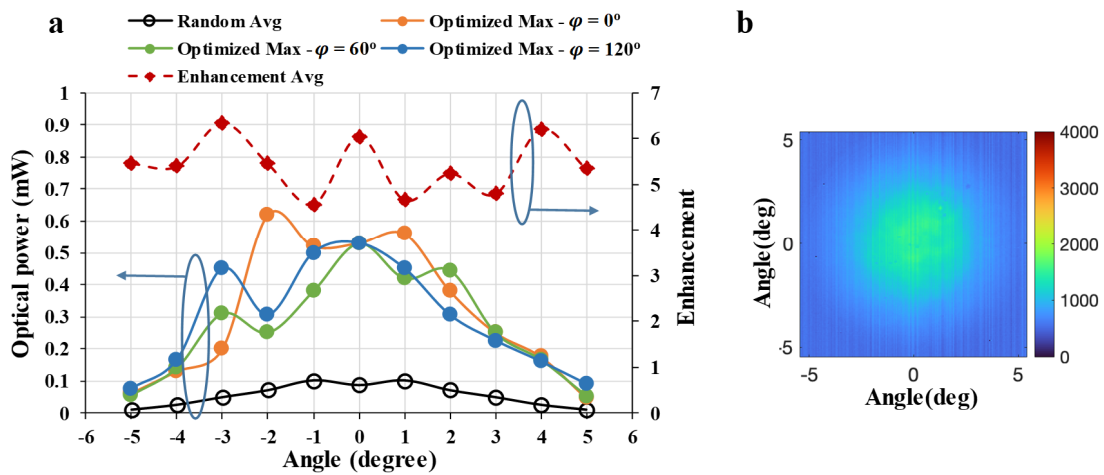

**Fig. S2** (a) Received power by the 9.5-mm power sensor at different steering angles $\eta$ and azimuth angles $\varphi$. (b) CCD image of the receiver plane averaged over random paddle configurations.

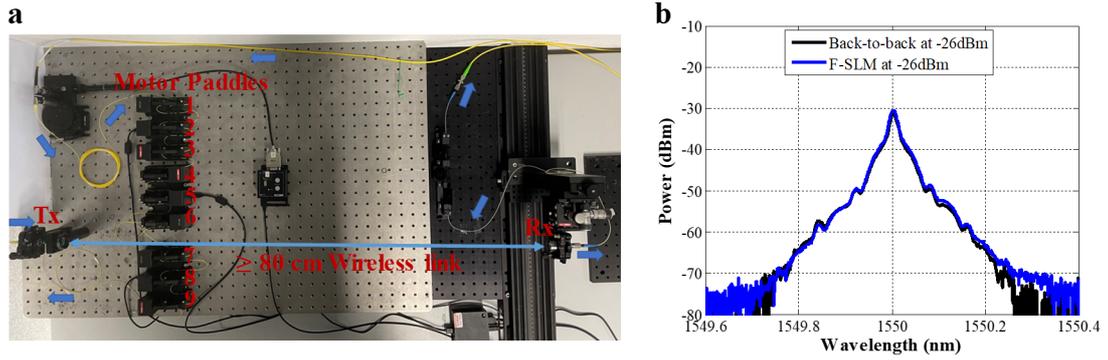

**Fig. S3** (a) A picture of the experimental setup. Tx: transmitter; Rx: receiver. (b) Measured back-to-back and F-SLM optical spectra of 10 Gb/s OOK signal at received optical powers of -26 dBm by an optical spectrum analyzer at the resolution of 0.02 nm.

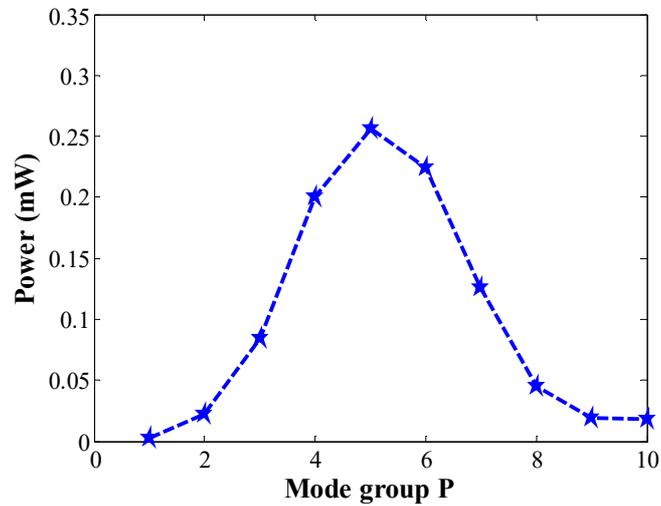

**Fig. S4** Power distribution in each group with 15-μm offset launch. The input optical power is set to 1 mW.

6